\documentclass[10pt,journal]{IEEEtran}
\ifCLASSOPTIONcompsoc
  \usepackage[nocompress]{cite}
\else
  \usepackage{cite}
\fi
\ifCLASSINFOpdf
  \usepackage[pdftex]{graphicx}
\else
\fi
\usepackage{booktabs}

\usepackage{color}

\usepackage{amsmath}
\usepackage{url}

\begin{document}
%
\title{Utilising Low Complexity CNNs to Lift Non-Local Redundancies in Video Coding}

\author{Jan~P.~Klopp,~Liang-Gee~Chen,~and Shao-Yi~Chien\\National Taiwan University
\IEEEcompsocitemizethanks{
	\IEEEcompsocthanksitem This research was supported in part by the Ministry of Science and Technology of Taiwan (MOST 108-2633-E-002-001), National Taiwan University (NTU-108L104039), Intel Corporation, Delta Electronics (06HZA27001) and Compal Electronics.
	\IEEEcompsocthanksitem Jan P. Klopp and Liang-Gee Chen are with the DSP \& IC Design Lab at the Graduate Institute of Electrical Engineering, National Taiwan University, Taipei, Taiwan (correspondence to kloppjp@gmail.com).
	\IEEEcompsocthanksitem
	Shao-Yi Chien is with the Media IC \& System Lab at the Graduate Institue of Electronics Engineering, National Taiwan University, Taipei, Taiwan.
	\IEEEcompsocthanksitem Jan P. Klopp and Shao-Yi Chien are also associated with the National Taiwan University IoX Center.
	
}
}



\IEEEtitleabstractindextext{%
\begin{abstract}
	Digital media is ubiquitous and produced in ever-growing quantities. This necessitates a constant evolution of compression techniques, especially for video, in order to maintain efficient storage and transmission. In this work, we aim at exploiting non-local redundancies in video data that remain difficult to erase for conventional video codecs.\\
	We design convolutional neural networks with a particular emphasis on low memory and computational footprint. The parameters of those networks are trained on the fly, at encoding time, to predict the residual signal from the decoded video signal. After the training process has converged, the parameters are compressed and signalled as part of the code of the underlying video codec. The method can be applied to any existing video codec to increase coding gains while its low computational footprint allows for an application under resource-constrained conditions. Building on top of High Efficiency Video Coding, we achieve coding gains similar to those of pretrained denoising CNNs while only requiring about 1\% of their computational complexity.\\
	Through extensive experiments, we provide insights into the effectiveness of our network design decisions. In addition, we demonstrate that our algorithm delivers stable performance under conditions met in practical video compression: our algorithm performs without significant performance loss on very long random access segments (up to 256 frames) and with moderate performance drops can even be applied to single frames in high-resolution low delay settings.
\end{abstract}

}

\maketitle

\IEEEdisplaynontitleabstractindextext

\section{Introduction}
\IEEEPARstart{V}{ideo} streams make up the largest part of worldwide Internet traffic and still experience strong growth due to an increase in on demand video services as well as high resolution and high-frame-rate content. At the same time, video decoding needs to operate under real time requirements on mobile devices under computationally constrained conditions, demanding algorithms that can be efficiently implemented in hardware.

Though image and video compression have been long standing problems, they have only recently attracted broad attention from the computer vision and machine learning communities \cite{Wu2018,Minnen2018,Liu2019,Mentzer2018,Theis2017,Toderici2016,Johnston2017,Rippel2017,Balle2017,Li2017,Baig2017,Toderici2015,Agustsson2017,Balle2018,Klopp}, bringing significant progress to the field of image compression. The increasing availability of highly efficient neural network accelerators renders machine learned solutions a viable alternative due to their easy adaptability to different data distributions and their relative independence from special purpose hardware.

Most existing and widely applied video compression techniques rely on two distinct mechanisms to exploit redundancies: first, motion compensation is used for content that can be reached through spatio-temporal references and, second, (residual) image coding for content not yet available or cannot be predicted by the decoder. Traditional encoding techniques perform both in a block-wise scheme where each block carries information about its temporal or spatial reference of an adjacent and already decoded block as well as about residual. More advanced coding techniques use variable block sizes and predict adjacent blocks, however, they still exploit redundancies that are temporally and spatially local. The same holds for recent machine-learning based approaches. They have replaced block-wise processing by deep convolutional neural networks. The generated code, however, is still local, bound to a certain position in the image, making it difficult to exploit global statistics.

Our approach is to utilize the ability of convolutional neural networks to compactly represent complex mappings inferred from large amounts of data in order to exploit non-local redundancies. In our case, the neural network's parameters are part of the code, hence the code is not tied to a certain part of the data. More concretely, we employ a CNN to predict the residual error of an existing encoder. The CNN thereby needs to be fit only to a particular segment of a video sequence instead of generalising to all possible videos sequences, which significantly reduces its computational footprint. Where the encoder only exploits local spatial or temporal redundancies, the neural network is optimised over an entire group of pictures at a time, thereby lifting non-local redundancies.

In summary, our contributions are as follows:
\begin{itemize}
	\item We introduce a network structure that is lightweight enough to be trained on the fly and signalled to the decoder and can still achieve coding gains of up to 6.8\% and 9.1\% in random access and low delay mode, respectively, for luma. Chroma gains rise up to 15.2\% and 19.5\% respectively.
	\item An algorithm for stably training and compressing the neural network on the fly without further hyper parameter search is presented.
	\item We evaluate our approach and elements of our algorithm design on the HEVC test sets and compare to a pretrained CNN method.
\end{itemize}

The remainder of our paper is structured as follows. Section~\ref{sec:related_work} introduces works related to our method. Section~\ref{sec:our_method} motivates and describes our approach and we report experimental results in Section~\ref{sec:experiments}. Section~\ref{sec:complexity} analyses and compares the complexity of our proposed algorithm. Finally, Section~\ref{sec:conclusion} concludes the paper and gives and outlook on future research in this direction.

\section{Related Work}\label{sec:related_work}
Our approach bears resemblance to conventional signal denoising filters. Such denoising filters can be found in recent video codecs, such as H.264 (AVC) \cite{Wiegand2003} or H.265 (HEVC) \cite{Sullivan2012a}. In their simplest form, deblocking filters \cite{Kim1999,List2003,Norkin2012,Jo2016} are employed to remove artifacts at the block boundaries. More recently, sample adaptive offset filtering \cite{Fu2012} has been developed as part of HEVC, which targets not only block boundaries but all pixels within a block. Even more flexible is adaptive loop filtering (ALF) \cite{Tsai2013a}, which exploits Wiener filter theory to derive an optimal linear denoising operator. This happens at encoding time with respect to an entire slice or a single block. The resulting filter parameters are then explicitly signalled to the decoder. In a more advanced version, \cite{Zhang2017} proposed a non-local ALF that represents the noise-free signal as a low rank approximation of patches of the decoded signal. While increasing coding gains, their method is more costly, especially for the decoder, as it relies on singular value decomposition. Krutz et al. \cite{Krutz2012} took a different direction and derived optimal filtering for multiple frames under motion estimation errors. These approaches can be seen as simpler predecessors to the proposed algorithm. While they face less challenges from signaling overhead or computational complexity due to their simpler nature, this also limits their gains. Furthermore, they often model linear dependencies while a neural network extends to more complex non-linear function, reaching improvements that linear filters cannot realize.

Recently, several methods based on convolutional neural networks have been proposed. Dong et al. \cite{Dong2015} introduced a CNN based method to suppress JPEG compression artefacts after decoding. \cite{Zhang2017b,Zhang2017a,Zhang2017c} propose a CNN-based image prior for denoising, enabling "blind" denoising without assumptions over the noise distribution. Yan et al. \cite{Yan2017} introduced a frame interpolation neural network that interpolates motion estimates to sub-pixel accuracy, improving coding gain through better motion vectors. Several works \cite{Yang2018,Li2017a,Yang2017,Cavigelli2017,Dai2017} employ CNNs to denoise HEVC compressed frames, where they distinguish between different slice types and quantisation levels. \cite{Zhang2018} explored the residual network architecture for this task, Jia et al. \cite{Jia2019} showed that ensembles yield further improvements.

In a different direction, the machine learning community has recently taken on the problem of image compression. Early approaches \cite{Toderici2015,Toderici2016,Johnston2017} adopted a residual encoding approach with recurrent neural networks. Later approaches \cite{Rippel2017,Balle2017} took the variational autoencoder as a basis and augmented it with code length regularisation, thereby reaching shorter codes at less complexity. Their algorithms were extended by context models \cite{Mentzer2018,Balle2018,Minnen2018,Klopp,Liu2019} to generate hierarchical codes, vector quantisation \cite{Agustsson2017}, content weighting \cite{Li2017} to control which parts of the image receive more bits and inpainting \cite{Baig2017} which predicts adjacent patches in the image space.

Finally, based on the aforementioned results from CNN-based image coding, several works \cite{Wu2018,Lu2019,Rippel2018} have proposed to perform motion and residual coding with neural networks. While their approaches are promising, they do not achieve results comparable to HEVC and are computationally expensive. All CNN-based methods in the literature that we are aware of share this high complexity characteristic. In contrast, a key feature of our algorithm is the adoption of an advanced machine learning model at low complexity and this sets our approach apart from other machine learning based algorithms.
\begin{figure*}[ht!]
	\centering
	\includegraphics[width=\textwidth]{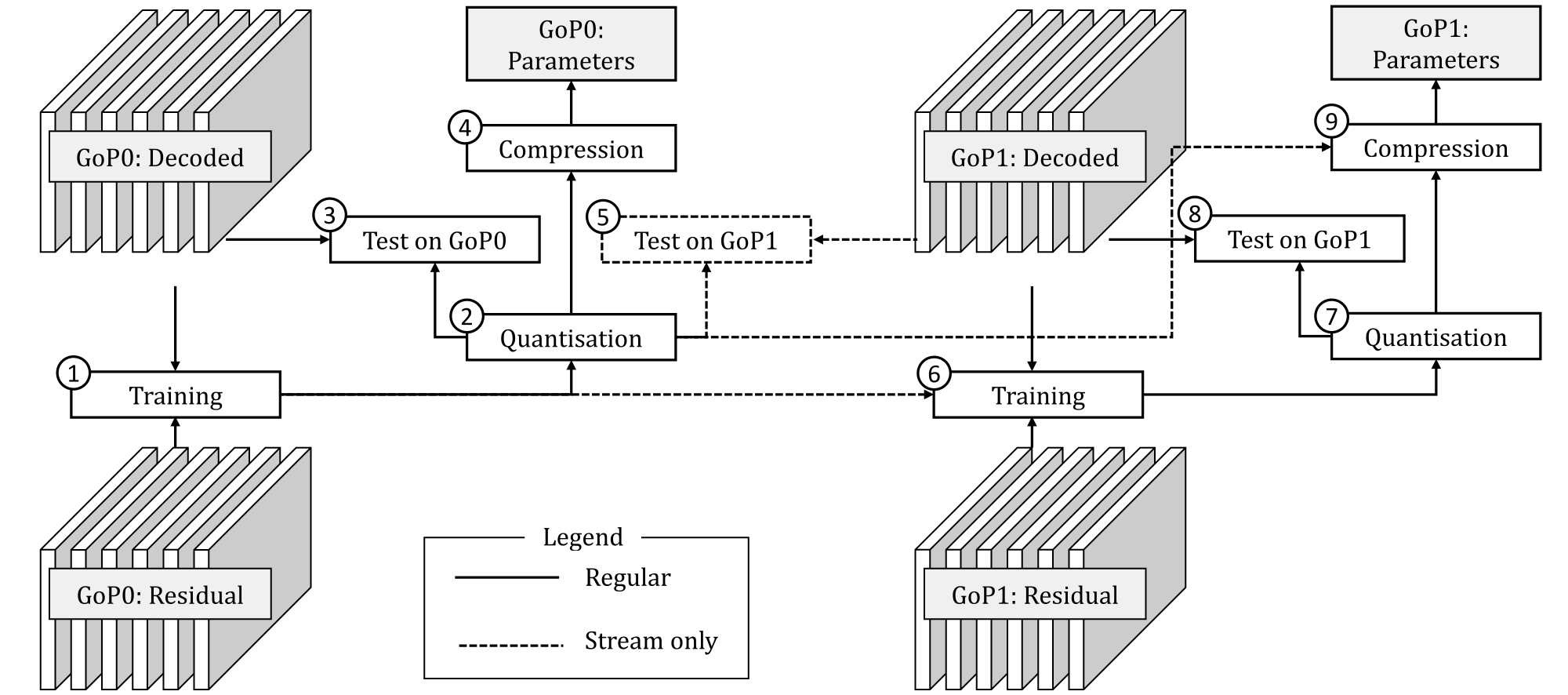}
	\caption{Encoding Process. The video signal is split into groups of pictures (GoP), the residuals of which are jointly predicted by a CNN that is trained on the fly (1). The CNN parameters are quantised (2) and the resulting CNN is tested (3) for coding gains on the GoP. If the test is positive, its parameters are compressed (4) before they are added to the bit stream of the underlying video codec. The dashed arrows/boxes indicate data transfer/operations that are only carried out in streaming scenarios where access to data signalled for previous frames is granted at the decoder. In such a streaming scenario, previously signalled parameters are first tested on the following GoP (5), before fine-tuning on that GoP (6) commences. Quantisation of those fine-tuned parameters (7) is followed by another test (8) to compare if higher gains can be achieved. If this is the case, the difference between new and old parameters is compressed (9) and added to the bit stream.}
	\label{fig:system_online}
\end{figure*}
\section{Exploiting Non-Local Redundancies}\label{sec:our_method}
Conventional video compression exploits redundancies between temporally and spatially adjacent parts in video sequences. With higher resolutions and more details in video footage, exploiting non-local redundancies becomes harder: conventional codecs would need to either increase their block sizes to capture patterns in a single set of transform coefficients or search across a larger set of blocks. Larger blocks, however, have less homogeneous content, making it difficult to capture structure in the image with a few quantised transform coefficients and a larger block search range in both spatial and the temporal direction leads to cubic complexity growths. At lower bit rates, these insufficiencies increase the chance of artefacts being present in the image if bandwidth does not allow fine grained quantisation. Deblocking filters \cite{Jo2016,Kim1999,List2003,Norkin2012}, sample adaptive offset \cite{Fu2012} or adaptive loop filtering \cite{Zhang2017,Tsai2013a} have been developed to counter those artefacts. These techniques, however, are applied locally, to a single block or a single slice, making it difficult to lift redundancies that are distributed across the temporal axis. Furthermore, only linear functions are used, limiting the expressiveness of the artefact suppression.

Our approach is to encode those parts of the residual that are redundant, i.e. details that are non local. These details need not necessarily originate from the same texture or pattern but need to have a similarity that can be encoded into a neural network conditioned on the decoded signal. We train a CNN on the fly at encoding time to predict the residual signal from the decoded one. As will be shown, the major advantage is that one can achieve coding gains with less operations compared to pretrained deep learning based denoising or loop filtering approaches. To achieve this, the network needs to be small and converge fast enough to make online training computationally feasible. At the same time, the network's parameters need to have a short code length so that their overhead on the existing bit stream does not cannibalise coding gains. In the following, we describe how to meet these challenges to yield an algorithm that can operate in an AutoML fashion to train the network. An overview of the algorithm is shown in Figure~\ref{fig:system_online}. Our algorithm can work in random access as well as low delay settings. The major difference is that in the low delay setting, the CNN's parameters can reference parameters previously signalled, thereby achieving lower compression rates. Besides this, our algorithm proceeds in the same way for both streaming and non-streaming video data. Its input is the decoded data from an existing video codec as well as the residual to be predicted. After training, the parameters are quantised and the network is run with the quantised parameters to test for an improvement, i.e. reduction of the residual error. If the test is successfull, the parameters are compressed and signalled. In a streaming scenario, this test is repeated on the next group of pictures so that after fine-tuning on that group, a new set of parameters is only signalled if the PSNR improvement of the new parameters exceeds 110\% of that of the previous parameters, thereby saving additional overhead. In the following, more details on the choice of network architecture, the parameter compression and the optimisation are given.
\subsection{Network Architecture}
\begin{table*}[ht!]
	\centering
	\caption{Network Architecture and Complexity for the Y Channel. Details of the five layer and 12 filter network architecture used for most experiments. Complexities are given in multiply-accumulate (MAC) operations per pixel for different pixel packaging configurations.}
	\label{tbl:network_architecture_y}
	\begin{tabular}{@{}lrrrrrrr@{}}
		\toprule
		Layer & Channels & Filters & Kernel & \multicolumn{4}{c}{Complexity (MAC/Pixel)} \\ 
		& & & & $P_H=1$, $P_W=1$ & $P_H=1$, $P_W=2$  & $P_H=2$, $P_W=1$  & $P_H=2$, $P_W=2$  \\ \midrule
		1      & $P_H\cdot P_W$  & 12      & $1\times 1$  & 12 & 12 & 12 & 12 \\
		2       & 1      & 12       & $3\times 3$  & 108 & 54 & 54 & 27   \\
		3       & 12      & 12       & $1\times 1$  & 144 & 72 & 72 & 36   \\
		4     	 & 1     & 12     & $3\times 3$ & 108 & 54 & 54 & 27  \\
		5    	 & 12   & $P_H\cdot P_W$       & $1\times 1$ & 12 & 12 & 12 & 12   \\
		\midrule
		Total       & &  & & 384 & 204 & 204 & 114 \\ \midrule
		& & & & \multicolumn{4}{c}{Number of Parameters}\\ \midrule
		Weights & & & & 384 & 408 & 408 & 456 \\
		Bias & & & & 48 & 48 & 48 & 48\\
		  \bottomrule
	\end{tabular}
\end{table*}
\begin{table*}[ht!]
	\centering
	\caption{Network Architecture and Complexity for the concatenated UV channels. Details of the five layer and 12 filter network architecture used for most experiments. Complexities are given in multiply-accumulate (MAC) operations per pixel for different pixel packaging configurations under the assumption that U and V channels are subsampled as in "YUV420".}
	\label{tbl:network_architecture_uv}
	\begin{tabular}{@{}lrrrrrrr@{}}
		\toprule
		Layer & Channels & Filters & Kernel & \multicolumn{4}{c}{Complexity (MAC/Pixel)} \\ 
		& & & & $P_H=1$, $P_W=1$ & $P_H=1$, $P_W=2$  & $P_H=2$, $P_W=1$  & $P_H=2$, $P_W=2$  \\ \midrule
		1      & $P_H\cdot P_W\cdot 2$  & 12      & $1\times 1$  & 6 & 6 & 6 & 6 \\
		2       & 1      & 12       & $3\times 3$  & 27 & 13.5 & 13.5 & 6.75   \\
		3       & 12      & 12       & $1\times 1$  & 36 & 18 & 18 & 9   \\
		4     	 & 1     & 12     & $3\times 3$ & 27 & 13.5 & 13.5  & 6.75  \\
		5    	 & 12   & $P_H\cdot P_W\cdot 2$       & $1\times 1$ & 6 & 6 & 6 & 6   \\
		\midrule
		Total       & &  & & 102 & 57 & 57 & 34.5 \\\midrule
		& & & & \multicolumn{4}{c}{Number of Parameters}\\ \midrule
		Weights & & & & 408 & 456 & 456 & 552 \\
		Bias & & & & 48 & 48 & 48 & 48\\  \bottomrule
	\end{tabular}
\end{table*}
The network architecture needs to be expressive enough to correct noise in the input video stream and at the same time lightweight enough to maintain a low computational footprint and a low signalling overhead when it's compressed and sent to the decoder. For this reason we chose an architecture inspired by MobileNets\cite{Howard2017}. MobileNets have been shown to work well in image recognition tasks and they combine the expressive power of deep neural networks with low computational and parameter size complexity. The basic idea of MobileNets is to factorise the convolutional layers, which are determined by their filters of dimensions $F\times C\times K_H\times K_W$ with $F$ feature maps, $C$ input channels and kernel height and width given by $K_H$ and $K_W$, respectively. Two separate convolutional layers are used to represent the same function, one operating only in the spatial domain with an independent filter for each input channel ($C\times 1\times K_H\times K_W$), and the other only connecting different channels with a $1\times 1$ kernel ($F\times C\times 1 \times 1$). \\
\begin{figure}[h!]
	\centering
	\includegraphics[width=\linewidth]{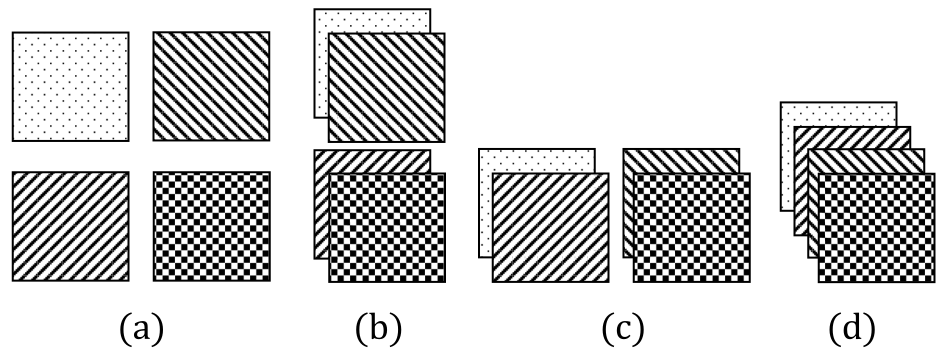}
	\caption{Pixel packing. Every square represents a single pixel. Patches of pixels are reorganised into vectors, which get treated like different channels by the CNN. $P_H/P_W$ denote height and width of a patch: $1/1$ equals no pixel packing (a), $1/2$ is shown in (b), $2/1$ in (c) and $2/2$ in (d).}
	\label{fig:pixel_packing}
\end{figure}
However, even with these techniques, the network's complexity may still be too high for high resolution content. For further reduction, we take inspirations from video codec design. Newer codecs like H.265 or H.264 profit from larger coding unit sizes as shown by Ohm et al. in \cite{Ohm2012}, in particular for higher resolutions. We exploit the fact that higher resolution videos have more homogeneous areas to reduce the complexity of our approach even further. We use pixel packing (Figure~\ref{fig:pixel_packing}) where a patch sized $P_H\times P_W$ of the input image is rearranged to a vector with $P_H\cdot P_W$ elements. This way, several pixels are processed within the same convolution, hence reducing the pixel-wise complexity by a factor $\frac{1}{P_H\cdot P_W}$. At the same time, the receptive field is enlarged without additional layers or layers with spatially larger filters. The network predicts a vector of $P_H\cdot P_W$ elements that are rearranged to form the residual prediction in the same shape as the input.

Our approach relies on optimising a non-convex function at encoding time using stochastic gradient descent. Unlike for convex optimisation, convergence guarantees for this non-convex problem are harder to obtain, if at all. Batch Normalization (BN)\cite{Ioffe2015} has been shown to greatly improve convergence behaviour of deep neural networks. We apply a Batch Normalization layer before each convolutional layer, except the first. Lastly, we remove the bias from the last convolutional layer as the overall residual we are predicting is bias-free. A single batch, however, may have a bias, yet this is what should be predicted from the input data instead of falsely adapting a fixed bias added to the prediction of the network. 

With the network architecture considerations presented above, we can use three hyper parameters to adjust the network complexity: the number of layers, the number of channels and pixel packaging. We found that a simple network of five layers and 12 channels works well while still guaranteeing low computational footprint. In addition, such a small network allows for efficient hardware implementation where all layers are processed jointly without intermediate DRAM memory access as shown in \cite{Chih2018}, making hardware realisations simpler. Table~\ref{tbl:network_architecture_y} lists each layer along with the pixel-wise complexity with and without different pixel packing choices. The chosen network requires from 114 to 384 operations per pixel for the Y channel. The U and V channels are processed jointly by a single pass through one network. The two chroma channels are concatenated, increasing the number of input channels/output filters of the network to $P_H\cdot P_W\cdot 2$. However, as the U and V channels in the popular "YUV420" format have only a quarter of the original pixels, the operations per pixel are lower as shown in Table~\ref{tbl:network_architecture_uv}.
Note that because luma and chroma are processed by different neural networks, their respective pixel packing configurations may differ.

\textcolor{black}{We found this network design to be beneficial when compared with vanilla convolutional neural networks. A three layer network with $3\times 3$ kernels and 12 channels performs about the same, however it has a complexity of 1514 and 408  MAC/Pixel for luma and chroma, respectively, when $P_H=P_W=1$. This is about four times the amount of our design with 384 and 102 MAC/Pixel.}
In total, our network design has a computational complexity low enough to allow real time dedicated hardware implementations for video compression at the decoder side even in mobile scenarios.

\subsection{Parameter Representation and Compression}
\label{sec:compression}
\textcolor{black}{During training time, parameters remain in 32 byte floating point format. From Tables \ref{tbl:network_architecture_y} and \ref{tbl:network_architecture_uv}, a the parameters take up 1728 and 1824 bytes for luma and chroma, respectively ($P_H=P_W=1$). Signalling this directly would void any distortion reductions in low bit rate cases. Hence, we present a scheme to efficiently represent and compress the parameters.}

Each layer (with the exception of the first) consists of a Batch Normalization (BN), a Convolution and a ReLU non-linearity. At encoding time, those are separated. After the optimisation routine has finished, the convolution and its batch normalisation layer can be merged into a single affine operation. The output $z$ at any position is given by 
\begin{equation}
\begin{split}
z&=\sum_c^{\text{Channel}}\sum_k^{\text{Kernel}} \frac{x_{c,k}-\mu_c}{\sigma_c}w_{c,k}+b\\
&=\sum_c^{\text{Channel}}\sum_k^{\text{Kernel}}(x_{c,k}-\mu_c)w'_{c,k}+b\\
&=\sum_c^{\text{Channel}}\sum_k^{\text{Kernel}}x_{c,k}w'_{c,k}+b'
\end{split}
\end{equation}
where $\mu_c$ and $\sigma_c$ are the BN parameters for channel $c$ and $w'_{c,k}$ is the weight for channel $c$ at kernel position $k$ that has been scaled by $\frac{1}{\sigma_c}$. As long as the size of the network parameters is small relative to the code length of the group of frames they accompany, there is no need for compression. However, in low latency live streaming scenarios, the neural network parameters may be updated and signalled every few frames, so that an efficient representation is necessary. This can be achieved by quantisation. While some approaches \cite{Courbariaux2016,Rastegari2016} quantise neural networks at training time to reduce the impact of quantisation, we did not find this beneficial in our experiments and as it adds additional overhead, we do quantise only once: after optimisation, before testing and signalling. Let $b_w$ and $b_b$ denote how many bits are used to quantise weights and bias, leading to quantisation ranges of $2^{b_w-1}-1$ and $2^{b_b-1}-1$, respectively. The weights are quantised by normalising them to the channel-wise quantisation range, $w^q_c=\lfloor0.5+\frac{w'_c}{\alpha_c}\rfloor$, where the scaling factor $\alpha_c=\frac{2^{b_w-1}-1}{\max_k |w_k|}$ is signalled separately. Quantisation of the bias happens over all filters $f$ so that $b^q_f=\lfloor0.5+\frac{b'_f}{\beta}\rfloor$ where $\beta=\frac{2^{b_b-1}-1}{\max_f |b'_f|}$ is a signalled separately again. 

In a streaming scenario, the decoder can still access previously signalled information, i.e. from the last $n$ frames. Hence, we transfer only their change to the decoder, thereby reducing the code length required to signal the neural network parameters. The idea is to take the difference between quantised values of two time steps and use an arithmetic coder to compress the difference signal. For the bias $b^q$, this works well because it changes slowly. For the weights, most change originates from a change in the batch normalisation parameter $\sigma_c$. As these changes are captured by the scaling factor $\alpha_c$, the quantised weights $w^q$ are unaffected by this change, which significantly lowers the coding rate. The differences of the significand and the exponent of $\alpha_c$ and $\beta$ are encoded separately using the same method. 

The largest share in code length originates from the weights $w^q$. To reduce the code length of the neural network parameters in low bitrate settings, the parameters $\theta(t)$ for a group of pictures $t$ should not differ too much from those of the previous group, $\theta(t-1)$. This way, an arithmetic coder would find a very simple distribution to encode. We add a regularisation term $\mathcal{R}(\theta(t),\theta(t-1))$ to the loss function in order to control the differences between normalised weights $\bar{w}_{l,c,k}$ identified by layer $l$, channel $k$ and kernel position $k$ at time $t$, $\bar{w}_{l,c,k}(t)=\frac{w_{l,c,k}(t)}{\max_k w_{l,c,k}(t)}$. For layer $l$, the regularisation is then
\begin{equation}
	\mathcal{R}_l(\theta(t),\theta(t-1))=\sum_c^{C_l}\sum_k^{K_l}\left(\bar{w}_{l,c,k}(t)-\bar{w}_{l,c,k}(t-1)\right)^2
\end{equation}
The entire regularisation is 
\begin{equation}
\mathcal{R}(\theta(t),\theta(t-1))=\sum_l^\text{Layers L}\frac{1}{C_lK_l}\mathcal{R}_l(\theta(t),\theta(t-1))
\end{equation}
where we normalise by the number of channels, $C_l$ and kernel elements $K_l$ in layer $l$.
In practice, we give this term a weighting of 0.1 and add it to the $L_2$ reconstruction loss. Note that this is not applied in random access mode, as accessing previous weights is not possible at decoding time. Instead, weights are subjected to $L_2$ norm regularisation at optimisation time.

\subsection{Optimisation}
At encoding time, the neural network learns to minimise the squared residual error for a particular group of pictures. This process should converge quickly, especially in online low delay settings, and not demand extensive hyper parameter tuning. We implement several techniques to accomplish this goal. 

To counter instabilities during training, we normalise the loss by the average $L_1$ error of the group of pictures. This leads to a normalisation of the gradient given by the $L_2$ loss, i.e.
\begin{equation}
	\frac{\partial\mathcal{L}}{\partial \hat{y}}=\frac{1}{2}\frac{\hat{y}-y}{L_1(R)})
\end{equation} where $L_1(R)"$ is the average $L_1$ norm over each pixel of the residual. This way, the optimisation process will not become unstable or require different learning rates for sequences with a different MSE. An additional benefit is that $L_2$ weight regularisation and code length regularisation (see previous paragraph) weightings do no need to be adjusted for different magnitudes of the reconstruction loss.

To aid convergence, we apply Batch Normalisation (BN), which works by eliminating the mean of each channel and normalising its variance to 1 during training time while accumulating a "global" mean and variance to be used during inference. The accumulation process is usually realised by giving a momentum $\gamma$ to the accumulated estimate and changing it by the current estimate weighted by $1-\gamma$. For the estimated dataset mean $\hat{\mu}_c$ and the mean of the current batch, $\mu_c^t$:
\begin{equation}
\hat{\mu}_c^t = \gamma\hat{\mu}_c^{t-1} + (1-\gamma)\mu_c^{t}
\end{equation}
The momentum is often set to values around $\gamma=0.9$, which is fine for scenarios where many training iterations are performed. In our case, we choose $\gamma=0.3$ so that adoption can proceed much quicker. Adding to this that we prefer a small batch size for performance reasons, Batch Normalisation can become unstable because its normalisation alters the activation $x$ that is used to compute the weight update $\frac{\partial\mathcal{L}}{\partial w}=x\frac{\partial\mathcal{L}}{\partial z}$ where $\frac{\partial\mathcal{L}}{\partial z}$ is the back-propagated gradient and $w$ the convolution kernel weight. If, for example, the mean of one channel for a particular batch happens to be far from the dataset's mean, this may lead to a weight update that severely deteriorates overall performance. To counter this, we tie the estimated mean value during training to the accumulated mean value. One could derive a different BN equation where the global mean estimate forms a prior for the batch-wise mean estimation. However, this would render the standard BN implementation that hardware vendors provide in their libraries useless and lead to a slower custom implementation. Instead, at training time we simply pad the input of the network with two more lines of zeros on two sides (e.g. right and bottom) that are cropped after exiting the network. Thereby a part of the activations throughout the network are constants (there is a negligible diffusion from the $3\times 3$ kernels). Those constants vary little as long as the bias values are small, which is the case for our network. The constant values will cause the mean estimates to be pulled towards them, acting like a regularisation. 
Note that for the optimisation problem itself, there is no difference as the loss is computed over the cropped image.

\section{Experiments} \label{sec:experiments}
\begin{table*}[ht!]
	\centering
	\caption{Average BD Rate savings in Random Access mode for different pixel packings.}
	\label{tbl:exp_ra_packing}
	\begin{tabular}{@{}lrrrrrrrrrrrrrr@{}}
		\toprule
		& \multicolumn{4}{c}{Y Channel} & & \multicolumn{4}{c}{U Channel} & & \multicolumn{4}{c}{V Channel} \\ \midrule 
		$P_H/P_W$ & 1/1 & 1/2 & 2/1 & 2/2 & & 1/1 & 1/2 & 2/1 & 2/2 & & 1/1 & 1/2 & 2/1 & 2/2 \\ \midrule
		HEVC A & -5.6\% & -5.7\% & -5.9\% & \textbf{-6.8}\% & & \textbf{-13.0}\% & -10.8\% & -10.4\% & -9.0\% & & \textbf{-14.4}\% & -12.8\% & -12.8\% & -11.9\% \\
		HEVC B & -5.4\% & \textbf{-5.9}\% & -4.5\% & -4.4\% & & \textbf{-9.8}\% & -8.2\% & -7.4\% & -6.9\% & & \textbf{-10.8}\% & -8.8\% & -8.3\% & -7.8\% \\
		HEVC C & \textbf{-3.7}\% & -2.8\% & -2.6\% & -1.6\% & & \textbf{-9.4}\% & -6.5\% & -6.1\% & -4.8\% & & \textbf{-15.3}\% & -12.0\% & -11.4\% & -9.0\% \\
		HEVC D & \textbf{-3.4}\% & -2.2\% & -2.2\% & -0.8\% & & \textbf{-8.7}\% & -6.1\% & -5.8\% & -3.8\% & & \textbf{-14.5}\% & -10.5\% & -10.2\% & -7.3\% \\
		HEVC E & \textbf{-2.8}\% & -1.6\% & -2.2\% & 0.2\% & & \textbf{-5.9}\% & -3.2\% & -2.9\% & -2.5\% & & \textbf{-10.5}\% & -7.1\% & -7.0\% & -7.1\% \\
		\bottomrule
	\end{tabular}
\end{table*}
\subsection{Setup}
We implement our approach in PyTorch 1.0 \cite{Paszke2017} and run our experiments on an Nvidia 1080 GPU. CuDNN's benchmark mode is disabled, its determinism enabled. As outlined in the previous section, our method needs to automatically apply to any sequence in any dataset and therefore we do not apply any sequence-wise or dataset-wise tuning. Albeit such hyper parameter optimisation being an active topic in research \cite{NIPS2012_4522}, it is to date only feasible in large scale operations. 

All our experiments use Adam \cite{Kingma2015} as optimiser and a learning rate of 0.02. During training we randomly sample non-overlapping patches of size $48\times 48$ and form batches of 64 patches. The weights are left to the standard PyTorch initialisation procedure, the bias is explicitly set to 0. These hyperparameters are the same for each sequence in each test set.

Our experiments are based on the HEVC Test Model HM-16.17, \textcolor{black}{inclding SAO and DB filters. Following the HEVC Common Test Conditions (CTC),} we evaluate our approach on the HEVC test sequences A to E in random access and low delay P and B modes. Our experimental results are reported separately for each of the three settings. We use the BD Rate \cite{Bjontegaard2001} savings to HM-16.17 to measure performance. For each channel, the rate savings are computed based on the channel's PSNR value and the total rate, i.e. the rate after the CNN filter has been applied to all channels. This is in accordance with video coding standards.

\textcolor{black}{As described in Section~\ref{sec:compression}, we quantise both weights and bias to have $b_w$ and $b_b$ bits, respectively. The number of bits are a trade-off between the accuracy and the signalling overhead. For the bias bit-width we found that $b_b=10$ worked well across all QPs and compression modes. To find suitable bit-widths for the weights without overfitting to specific sequences or coding modes, we used non-CTC sequences \cite{Derf2020} to determine the optimal number of weight bits. We used several sequences of different resolutions in the low delay B setting, as it is the most challenging one, due to the small GoP size and bi-directional prediction. We found that 10, 9, 7, 6 bits worked best for QPs 22, 27, 32, and 37, respectively, and used the those quantizations throughout all experiments.}

\textcolor{black}{While our approach is suitable for both in-loop and ex-loop application, our experiments make use of the ex-loop variant. This has the advantage of not requiring two encoding passes, where the first is used to generate the training data and the second one applies the trained neural network. Thereby, our approach increases encoding times only moderately for large resolutions.}
\subsection{Random Access}
In the random access scenario, the network parameters for a particular random access segment (RAS) are independent of those belonging to previous or later segments. In practice, RAS are often encoded in parallel as this provides a linear speedup. To be compatible to this approach, we learn network parameters for each RAS from scratch. 

We present several series of experiments to analyse different aspects of the proposed algorithm. At first, we analyse the influence of pixel packing on the optimisation performance. As described in Section~\ref{sec:our_method}, pixel packing increases the receptive field size of the model, but requires joint prediction of several pixels at the same time. Table~\ref{tbl:exp_ra_packing} shows BD Rate savings for different pixel packings for each HEVC test set and channel. For the Y channel, high resolution tends to benefit from a larger receptive field even if the complexity per pixel is reduced, for HEVC A a $2/2$ packing yields significantly higher results than all other variants, HEVC B peaks for $1/2$, where two horizontally adjacent pixels are packed. Performance on smaller resolutions like HEVC C and D on the other hand almost halves when $2/2$ pixel packing is applied, the reason for this may lie in the higher information density of a low resolution sequence, which does not benefit from a larger receptive field. For the chroma channels, the picture is clearer, all test sets peak when no pixel packing is applied. However, smaller resolutions suffer higher losses when packing is applied. Overall, this shows that pixel packing helps where it is needed most: in high resolutions where the significantly lower number of operations per pixel contributes most to reducing the overall cost for implementing this method.

Experiments conforming with the HEVC Test Model are constrained to use only up to 32 frames in random access mode to guarantee comparability to other methods. In practice, however, encoders utilise much longer intra frame periods, for example in online video streams where it's unlikely that the user will perform a lot of fine grained seek operations throughout the sequence. Applying the CNN to a larger set of frames at once has two advantages: 
\begin{itemize}
	\item The same amount of data is signalled for more frames, leading to less signalling overhead and thereby potentially to higher coding gains, especially for small resolutions.
	\item The same amount of computation is used. We observed that the convergence of the CNNs online training process is hardly influenced by the number of frames taken into the training data set. Hence, we leave the the number of patches per batch, the size of each patch and the number of training iterations constant.
\end{itemize}

\begin{table*}[ht!]
	\centering
	\caption{Average BD Rate savings in Random Access mode for different RAS segment sizes.}
	\label{tbl:exp_ra_ras_size}
	\begin{tabular}{@{}lrrrrrrrrrrrrrr@{}}
		\toprule
		& \multicolumn{4}{c}{Y Channel} & & \multicolumn{4}{c}{U Channel} & & \multicolumn{4}{c}{V Channel} \\ \midrule 
		\#Frames & 32 & 64 & 128 & 256 & & 32 & 64 & 128 & 256 & & 32 & 64 & 128 & 256 \\ \midrule
		HEVC A & -5.6\% & -6.0\% & -5.6\% & \textbf{-6.0}\% & & -13.0\% & -12.4\% & \textbf{-13.2}\% & -10.8\% & & -14.4\% & -13.8\% & \textbf{-15.2}\% & -12.4\% \\
		HEVC B & -5.4\% & \textbf{-5.7}\% & -5.3\% & -5.1\% & & -9.8\% & -9.4\% & \textbf{-10.0}\% & -9.3\% & & -10.8\% & -10.6\% & \textbf{-11.2}\% & -10.2\% \\
		HEVC C & -3.7\% & -4.1\% & \textbf{-4.2}\% & -4.1\% & & \textbf{-9.4}\% & -9.4\% & -9.4\% & -9.1\% & & -15.3\% & \textbf{-15.3}\% & -14.5\% & -14.1\% \\
		HEVC D & -3.4\% & -4.8\% & -5.4\% & \textbf{-5.5}\% & & -8.7\% & -8.8\% & \textbf{-9.7}\% & -9.4\% & & -14.5\% & -15.0\% & \textbf{-15.0}\% & -14.9\% \\
		HEVC E & -2.8\% & -3.9\% & -4.2\% & \textbf{-4.6}\% & & -5.9\% & -6.4\% & -7.4\% & \textbf{-7.9}\% & & -10.5\% & -11.0\% & -11.5\% & \textbf{-11.9}\% \\
		\bottomrule
	\end{tabular}	
\end{table*}

\begin{table*}[ht!]
	\centering
	\caption{Average BD Rate savings in Random Access mode for different complexities.}
	\label{tbl:exp_ra_comp}
	\begin{tabular}{@{}lrrrrrrrr@{}}
		\toprule
		& \multicolumn{2}{c}{Y Channel} & & \multicolumn{2}{c}{U Channel} & & \multicolumn{2}{c}{V Channel} \\ \midrule 
		$P_H/P_W$ & 2/2 & 1/1 & & 2/2 & 1/1 & & 2/2 & 1/1 \\ 
		\#Filters & 12 & 6 & & 12 & 6 & & 12 & 6 \\ 
		Complexity (MAC/Pixel) & 114 & 156 & & 34.5 & 42 & & 34.5 & 42 \\ \midrule
		HEVC A & \textbf{-6.8}\% & -4.7\% & & -9.0\% & \textbf{-9.3}\% & & \textbf{-11.9}\% & -11.0\% \\
		HEVC B & \textbf{-4.4}\% & -4.3\% & & \textbf{-6.9}\% & -6.4\% & & \textbf{-7.8}\% & -7.1\% \\
		HEVC C & -1.6\% & \textbf{-2.9}\% & & -4.8\% & \textbf{-5.7}\% & & -9.0\% & \textbf{-11.1}\% \\
		HEVC D & -0.8\% & \textbf{-3.4}\% & & -3.8\% & \textbf{-5.9}\% & & -7.3\% & \textbf{-11.2}\% \\
		HEVC E & 0.2\% & \textbf{-2.0}\% & & -2.5\% & \textbf{-4.0}\% & & -7.1\% & \textbf{-7.9}\% \\
		\bottomrule
	\end{tabular}
\end{table*}
Table~\ref{tbl:exp_ra_ras_size} shows experiment results for RAS lengths 32 (same as Table~\ref{tbl:exp_ra_packing}), 64, 128 and 256 for all channels and datasets. Note that we fixed pixel packing to $P_H=1,P_W=1$, however, the results should be equally applicable to other packing configurations. For the Y channel, across all datasets, we observe that despite increasing the number of pixels by 8-fold (i.e. from 32 to 256 frames), the performance drop is acceptable for HEVC B while average BD rate savings are increasing for all other test sets. The increase is most dramatic for HEVC D and E. HEVC D is small in size (416x240) while HEVC E features typical streaming content, similar to video conferencing. Hence, both test sets have low bit rates compared to the other sets. Because signalling the CNN's parameters adds an almost constant overhead to the bit stream, its negative effect on rate savings is most pronounced for low bit rate sequences. Signalling CNN parameters for more frames can then mitigate these effects if PSNR improvement is preserved as is the case for all test sets. For the chroma channels Table~\ref{tbl:exp_ra_ras_size} shows a similar pattern, albeit most sequences peak at 128 frames. Overall, this demonstrates that in practical applications where typically larger I frame periods are chosen, our approach can be used even in low bit rate conditions to achieve similar gains as for high bit rate sequences.

In the preceding section, we introduced pixel packing as a method to reduce the per pixel complexity of the neural network while maintaining its architecture. Table~\ref{tbl:exp_ra_comp} compares BD rate savings for the 2/2 pixel packed case with a downsized network architecture using only 6 filters in each layer. For higher resolutions (HEVC A \& B), pixel packing performs better or at least equal, despite having a lower complexity. At low resolutions, less filters per layer are the better option. This once more underlines that pixel packing's impact is largest where its complexity reduction is needed most.
\begin{table*}[ht!]
	\centering
	\caption{Average BD Rate savings in Low Delay B mode for different pixel packings at a GoP size of five frames.}
	\label{tbl:exp_ldb_packing}
	\begin{tabular}{@{}lrrrrrrrrrrrrrr@{}}
		\toprule
		& \multicolumn{4}{c}{Y Channel} & & \multicolumn{4}{c}{U Channel} & & \multicolumn{4}{c}{V Channel} \\ \midrule 
		$P_H/P_W$ & 1/1 & 1/2 & 2/1 & 2/2 & & 1/1 & 1/2 & 2/1 & 2/2 & & 1/1 & 1/2 & 2/1 & 2/2 \\ \midrule
		HEVC A & -5.2\% & -5.0\% & -5.1\% & \textbf{-5.9}\% & & \textbf{-15.5}\% & -13.3\% & -12.7\% & -11.8\% & & \textbf{-18.4}\% & -16.0\% & -15.5\% & -14.7\% \\
		HEVC B & -4.6\% & \textbf{-4.9}\% & -3.5\% & -3.4\% & & \textbf{-12.5}\% & -9.9\% & -9.3\% & -8.4\% & & \textbf{-16.8}\% & -13.1\% & -13.7\% & -12.0\% \\
		HEVC C & \textbf{-3.3}\% & -1.9\% & -1.7\% & 0.3\% & & \textbf{-12.1}\% & -8.6\% & -8.5\% & -5.7\% & & \textbf{-18.0}\% & -14.8\% & -14.5\% & -10.8\% \\
		HEVC D & \textbf{2.8}\% & 5.7\% & 5.7\% & 9.9\% & & \textbf{-7.8}\% & -2.5\% & -2.3\% & 2.8\% & & \textbf{-16.7}\% & -9.4\% & -10.0\% & -3.2\% \\
		HEVC E & \textbf{-3.4}\% & -1.3\% & -1.3\% & 3.1\% & & \textbf{-10.2}\% & -7.1\% & -5.7\% & -2.0\% & & \textbf{-17.0}\% & -15.3\% & -13.6\% & -10.7\% \\
		\bottomrule
	\end{tabular}
\end{table*}
\begin{table*}[ht!]
	\centering
	\caption{Average BD Rate savings in Low Delay P mode for different pixel packings at a GoP size of five frames.}
	\label{tbl:exp_ldp_packing}
	\begin{tabular}{@{}lrrrrrrrrrrrrrr@{}}
		\toprule
		& \multicolumn{4}{c}{Y Channel} & & \multicolumn{4}{c}{U Channel} & & \multicolumn{4}{c}{V Channel} \\ \midrule 
		$P_H/P_W$ & 1/1 & 1/2 & 2/1 & 2/2 & & 1/1 & 1/2 & 2/1 & 2/2 & & 1/1 & 1/2 & 2/1 & 2/2 \\ \midrule
		HEVC A & -9.0\% & -8.7\% & -8.8\% & \textbf{-9.1}\% & & \textbf{-15.7}\% & -13.9\% & -13.5\% & -11.6\% & & \textbf{-19.5}\% & -17.0\% & -17.3\% & -15.2\% \\
		HEVC B & -6.3\% & \textbf{-6.8}\% & -5.2\% & -5.3\% & & \textbf{-13.9}\% & -11.4\% & -10.8\% & -9.4\% & & \textbf{-17.8}\% & -15.1\% & -15.0\% & -12.7\% \\
		HEVC C & \textbf{-3.2}\% & -1.8\% & -1.7\% & 0.4\% & & \textbf{-12.4}\% & -8.4\% & -8.6\% & -5.8\% & & \textbf{-18.8}\% & -15.4\% & -14.9\% & -11.6\% \\
		HEVC D & \textbf{3.2}\% & 5.8\% & 6.0\% & 10.3\% & & \textbf{-7.9}\% & -2.6\% & -3.2\% & 2.1\% & & \textbf{-15.5}\% & -8.7\% & -11.1\% & -3.5\% \\
		HEVC E & \textbf{-5.3}\% & -3.0\% & -3.6\% & 1.8\% & & \textbf{-12.6}\% & -9.0\% & -7.5\% & -3.9\% & & \textbf{-19.7}\% & -15.9\% & -15.7\% & -11.3\% \\
		\bottomrule
	\end{tabular}
\end{table*}
\begin{table*}[ht!]
	\centering
	\caption{Average BD Rate savings in Low Delay B mode for different GoP lengths.}
	\label{tbl:exp_ldb_gop_size}
	\setlength\tabcolsep{4.5pt}
	\begin{tabular}{@{}lrrrrrrrrrrrrrrrrr@{}}
		\toprule
		& \multicolumn{5}{c}{Y Channel} & & \multicolumn{5}{c}{U Channel} & & \multicolumn{5}{c}{V Channel} \\ \midrule 
		\#Frames & 5 & 4 & 3 & 2 & 1 & & 5 & 4 & 3 & 2 & 1 & & 5 & 4 & 3 & 2 & 1 \\ \midrule
		HEVC A & \textbf{-5.2}\% & -5.0\% & -4.8\% & -4.6\% & -4.5\% & & \textbf{-15.5}\% & -15.2\% & -14.3\% & -14.2\% & -13.7\% & & \textbf{-18.4}\% & -18.0\% & -16.9\% & -17.1\% & -16.8\% \\
		HEVC B & \textbf{-4.6}\% & -4.3\% & -4.2\% & -4.0\% & -3.5\% & & \textbf{-12.5}\% & -12.0\% & -11.7\% & -11.0\% & -10.4\% & & \textbf{-16.8}\% & -16.2\% & -15.8\% & -15.1\% & -14.5\% \\
		HEVC C & \textbf{-3.3}\% & -2.7\% & -2.3\% & -1.4\% & 1.6\% & & \textbf{-12.1}\% & -11.6\% & -11.0\% & -10.1\% & -7.7\% & & \textbf{-18.0}\% & -18.0\% & -17.6\% & -16.7\% & -14.2\% \\
		HEVC D & \textbf{2.8}\% & 4.7\% & 6.9\% & 12.3\% & 28.3\% & & \textbf{-7.8}\% & -5.4\% & -3.2\% & 0.9\% & 14.5\% & & \textbf{-16.7}\% & -13.4\% & -11.6\% & -8.9\% & 3.6\% \\
		HEVC E & \textbf{-3.4}\% & -2.7\% & -2.0\% & 0.2\% & 6.2\% & & \textbf{-10.2}\% & -9.2\% & -9.3\% & -6.5\% & -0.6\% & & -17.0\% & \textbf{-17.5}\% & -16.4\% & -14.7\% & -8.4\% \\
		\bottomrule
	\end{tabular}
\end{table*}
\begin{table*}[ht!]
	\centering
	\caption{Average BD Rate savings in Low Delay P mode for different GoP lengths.}
	\label{tbl:exp_ldp_gop_size}
	\setlength\tabcolsep{4.5pt}
	\begin{tabular}{@{}lrrrrrrrrrrrrrrrrr@{}}
		\toprule
		& \multicolumn{5}{c}{Y Channel} & & \multicolumn{5}{c}{U Channel} & & \multicolumn{5}{c}{V Channel} \\ \midrule 
		\#Frames & 5 & 4 & 3 & 2 & 1 & & 5 & 4 & 3 & 2 & 1 & & 5 & 4 & 3 & 2 & 1 \\ \midrule
		HEVC A & \textbf{-9.0}\% & -8.6\% & -8.5\% & -8.4\% & -8.5\% & & \textbf{-15.7}\% & -15.0\% & -13.6\% & -14.3\% & -14.0\% & & \textbf{-19.5}\% & -18.4\% & -18.0\% & -18.4\% & -17.7\% \\
		HEVC B & \textbf{-6.3}\% & -6.2\% & -6.1\% & -5.9\% & -5.3\% & & \textbf{-13.9}\% & -13.5\% & -13.2\% & -12.6\% & -11.9\% & & \textbf{-17.8}\% & -17.4\% & -17.2\% & -16.5\% & -15.8\% \\
		HEVC C & \textbf{-3.2}\% & -2.8\% & -2.3\% & -1.1\% & 2.2\% & & \textbf{-12.4}\% & -11.4\% & -11.4\% & -9.7\% & -7.4\% & & -18.8\% & \textbf{-19.3}\% & -18.0\% & -17.1\% & -14.3\% \\
		HEVC D & \textbf{3.2}\% & 4.9\% & 7.7\% & 13.7\% & 28.9\% & & \textbf{-7.9}\% & -6.5\% & -3.6\% & 0.7\% & 13.9\% & & \textbf{-15.5}\% & -15.1\% & -13.2\% & -9.3\% & 2.4\% \\
		HEVC E & \textbf{-5.3}\% & -4.7\% & -3.8\% & -2.0\% & 4.2\% & & \textbf{-12.6}\% & -12.0\% & -11.2\% & -8.4\% & -2.3\% & & \textbf{-19.7}\% & -19.2\% & -18.9\% & -16.3\% & -9.5\% \\
		\bottomrule
	\end{tabular}	
\end{table*}
\begin{table*}[ht!]
	\centering
	\caption{Average BD Rate savings in Low Delay B mode for different complexities at a GoP size of two frames.}
	\label{tbl:exp_ldb_comp}
	\begin{tabular}{@{}lrrrrrrrr@{}}
		\toprule
		& \multicolumn{2}{c}{Y Channel} & & \multicolumn{2}{c}{U Channel} & & \multicolumn{2}{c}{V Channel} \\ \midrule 
		$P_H/P_W$ & 2/2 & 1/1 & & 2/2 & 1/1 & & 2/2 & 1/1 \\ 
		\#Filters & 12 & 6 & & 12 & 6 & & 12 & 6 \\ 
		Complexity (MAC/Pixel) & 114 & 156 & & 34.5 & 42 & & 34.5 & 42 \\ \midrule
		HEVC A & \textbf{-4.6}\% & -3.9\% & & \textbf{-14.2}\% & -11.0\% & & \textbf{-17.1}\% & -13.1\% \\
		HEVC B & \textbf{-4.0}\% & -3.1\% & & \textbf{-11.0}\% & -7.8\% & & \textbf{-15.1}\% & -11.2\% \\
		HEVC C & -1.4\% & \textbf{-2.4}\% & & \textbf{-10.1}\% & -7.5\% & & \textbf{-16.7}\% & -13.2\% \\
		HEVC D & 12.3\% & \textbf{2.4}\% & & 0.9\% & \textbf{-2.9}\% & & -8.9\% & \textbf{-9.9}\% \\
		HEVC E & 0.2\% & \textbf{-1.5}\% & & \textbf{-6.5}\% & -5.0\% & & \textbf{-14.7}\% & -13.0\% \\
		\bottomrule
	\end{tabular}
\end{table*}
\begin{table*}[ht!]
	\centering
	\caption{Average BD Rate savings in Low Delay P mode for different complexities at a GoP size of two frames.}
	\label{tbl:exp_ldp_comp}
	\begin{tabular}{@{}lrrrrrrrrr@{}}
		\toprule
		& \multicolumn{2}{c}{Y Channel} & & \multicolumn{2}{c}{U Channel} & & \multicolumn{2}{c}{V Channel} \\ \midrule 
		$P_H/P_W$ & 2/2 & 1/1 & & 2/2 & 1/1 & & 2/2 & 1/1 \\ 
		\#Filters & 12 & 6 & & 12 & 6 & & 12 & 6 \\ 
		Complexity (MAC/Pixel) & 114 & 156 & & 34.5 & 42 & & 34.5 & 42 \\ \midrule
		HEVC A & \textbf{-8.4}\% & -7.6\% & & \textbf{-14.3}\% & -10.8\% & & \textbf{-18.4}\% & -13.4\% \\
		HEVC B & \textbf{-5.9}\% & -5.3\% & & \textbf{-12.6}\% & -9.8\% & & \textbf{-16.5}\% & -12.5\% \\
		HEVC C & -1.1\% & \textbf{-2.2}\% & & \textbf{-9.7}\% & -7.6\% & & \textbf{-17.1}\% & -13.7\% \\
		HEVC D & 13.7\% & \textbf{3.1}\% & & 0.7\% & \textbf{-3.0}\% & & -9.3\% & \textbf{-10.8}\% \\
		HEVC E & -2.0\% & \textbf{-2.9}\% & & \textbf{-8.4}\% & -6.4\% & & \textbf{-16.3}\% & -14.4\% \\
		\bottomrule
	\end{tabular}
\end{table*}
\subsection{Low Delay B/P}
The low delay setting is more challenging for our approach compared to the random access setting as the GoP size is reduced to only a few frames. The signalled parameters are likely to cause a higher bit rate overhead in this scenario. On the other hand, data that is already available at the decoder can be reused. Therefore, for our experiments, we signal new parameters only if their PSNR improvement is more than 10\% higher than what the previously signalled CNN would achieve when applied to unseen frames of the following GoP. In practice, this enables our algorithm to be applied to lower resolution settings even for small GoP sizes.

Following our analysis in the random access setting, we first evaluate the influence of pixel packing on the different test sets for both LD B and P settings. The results are listed in Tables~\ref{tbl:exp_ldb_packing} and \ref{tbl:exp_ldp_packing}, and reflect the preference of higher resolutions for larger receptive fields through pixel packing as observed in the random access setting for the Y channel. For chroma channels, no pixel packing yields the best results, however, for large resolutions the performance drop is much smaller than for lower bit rates. As mentioned in the previous paragraph, the small number of frames per GoP makes it challenging to apply the algorithm to low bit rate sequences. While there are still significant BD rate savings for HEVC C and E, the algorithm fails when applied to HEVC D, as seen from the positive rate savings in both low delay variants. In addition, application to the LDP mode yields greater improvement. LDP allows only prediction in one direction and hence gives the encoder less options to optimise the code and reduce the residual. It is hence plausible that the CNN based residual prediction has a higher chance of lifting unexploited patterns in the residual signal.

With a GoP size of five frames, we follow a common setting. However, in some applications an even lower latency is favourable. Tables~\ref{tbl:exp_ldb_gop_size} and \ref{tbl:exp_ldp_gop_size} show BD rate savings for different GoP sizes, down to a single frame, for LDB and LDP, respectively. Unsurprisingly, a larger GoP size performs best in all settings. It is evident, though, that the proposed algorithm can achieve significant rate savings even if only a single frame is processed at a time for the high resolution sequences in HEVC A \& B. For smaller resolutions, HEVC C \& E, chroma channels still hold gains if the GoP size is reduced, however in the luma domain rate savings turn positive as initial coding gains did not reach similar levels to gains in the chroma domain. In the smallest resolution, HEVC D, gains vanish even for U and V channels if the algorithm runs in single frame mode as the signalling overhead cannibalises any gains achieved by the neural network. 

To measure the efficacy of pixel packing as complexity reduction, Tables~\ref{tbl:exp_ldb_comp} and \ref{tbl:exp_ldp_comp} compare 2/2 pixel packaging to downsizing by reducing the number of filters per layer. Similar to the random access results before, pixel packing is an efficient option for high resolution in the Y channel. For chroma channels, there is a significant drop in performance despite higher complexity when trading pixel packing for less filters. Overall, this underlines the importance of choosing the right complexity reduction method.

\begin{figure*}[h]
	\centering
	\includegraphics[width=1.0\linewidth]{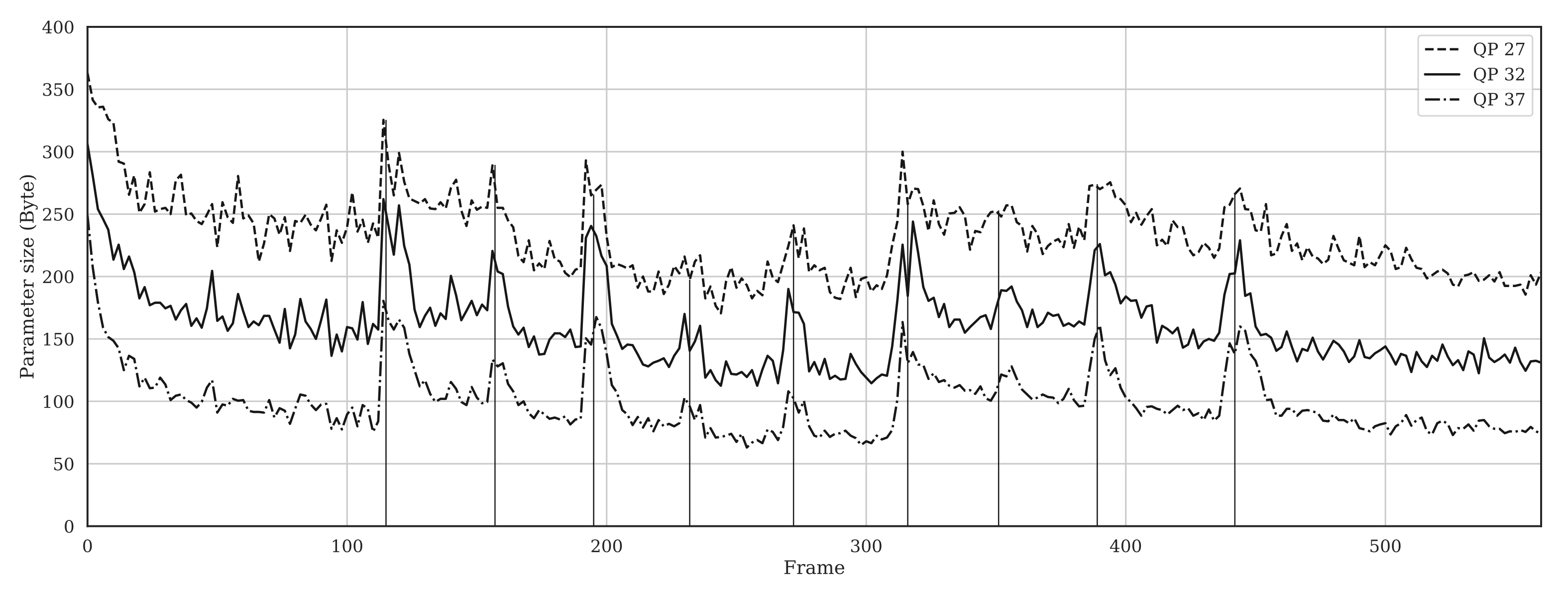}
	\caption{Overhead per frame caused by signalling the CNN parameters over time for different QP values when applied to the "Controlled Burn" test sequence luma channel with GoP length of two. Vertical lines indicate scene changes.}
	\label{fig:controlledburnbitrate}
\end{figure*}

Finally, Fig.~\ref{fig:controlledburnbitrate} shows the progression of frame wise signalling overhead for the test sequence "Controlled Burn" \textcolor{black}{for the luma network at a GoP length of two}. The sequence features several scene changes and fast forwards that are indicated by vertical lines. We've plotted graphs for three different QPs. Note that "skipping" a network, i.e. indicating that the previously signalled network ought to be reused, is deactivated to emphasise the data rate behaviour in this case. The graphs reside at different levels as the weights are quantised more coarse for higher QPs. \textcolor{black}{Compared to the size of 1728 bytes that the uncompressed floating point parameters take (see Section~\ref{sec:compression}), the sizes of the initial network range from 500 to 740 bytes (frame-wise size multiplied by GoP length). The compression rate is about 28.9\% (QP 37) to 42.8\% (QP 27) for a network encoded without referencing previously signalled parameters. It is lowered significantly when scene statistics allow differential encoding of parameters based on the previous GoP's network parameters.}
The abrupt changes in scene statistics cause a small upturn in the bitrate that is quickly reduced after a few frames. The first value (for frame 0) indicates the code size without reference to prior network parameters. It can easily be observed that despite scene changes, the code size is regularly settling far below that initial value. This shows that our modified loss function together with difference coding work effectively to yield a significant code size reduction in low delay stream settings.
\begin{table*}[ht]
	\centering
	\caption{Comparison of average BD Rate savings and complexity with a pretrained CNN approach in Random Access mode.}
	\label{tbl:exp_ra_comparison}
	\begin{tabular}{@{}lrrrrrrrr@{}}
		\toprule
		& \multicolumn{4}{c}{Our's}                                                                                  & \multicolumn{4}{c}{Jia et al. \cite{Jia2019}}                                                            \\ \midrule
		& \begin{tabular}[c]{@{}r@{}}Complexity (Y+U/V)\\ (MAC/Pix)\end{tabular} & Y      & U       & V       & \begin{tabular}[c]{@{}r@{}}Complexity\\ (MAC/Pix)\end{tabular} & Y      & U      & V      \\ \midrule
		HEVC A & 216 (114+102) & \textbf{-6.8}\% & \textbf{-13.0}\% & \textbf{-14.4}\% & 326336  & -6.6\% & -3.4\% & -3.0\% \\
		HEVC B & 306 (204+102) & -5.9\% & \textbf{-9.8}\%  & \textbf{-10.8}\% & 326336  & \textbf{-6.5}\% & -2.5\% & -2.7\% \\
		HEVC C & 486 (384+102) & -3.7\% & \textbf{-9.4}\%  & \textbf{-15.3}\% & 326336  & \textbf{-4.5}\% & -3.3\% & -4.5\% \\
		HEVC D & 486 (384+102) & \textbf{-3.4}\% & \textbf{-8.7}\%  & \textbf{-14.5}\% & 326336  & -3.3\% & -2.6\% & -3.6\% \\
		HEVC E & 486 (384+102) & -2.8\% & \textbf{-5.9}\%  & \textbf{-10.5}\% & 326336  & \textbf{-9.0}\% & -4.2\% & -5.3\% \\ \bottomrule
	\end{tabular}
\end{table*}
\begin{table*}[ht]
	\centering
	\caption{Comparison of average BD Rate savings and complexity with a pretrained CNN approach in Low Delay B mode.}
	\label{tab:exp_ldb_comparison}
	\begin{tabular}{@{}lrrrrrrrr@{}}
		\toprule
		& \multicolumn{4}{c}{Our's} & \multicolumn{4}{c}{Jia et al. \cite{Jia2019}} \\ \midrule
		& \begin{tabular}[c]{@{}r@{}}Complexity (Y+U/V)\\ (MAC/Pix)\end{tabular} & Y & U & V & \begin{tabular}[c]{@{}r@{}}Complexity\\ (MAC/Pix)\end{tabular} & Y & U & V \\ \midrule
		HEVC A & 216 (114+102) & -5.9\% & \textbf{-15.5}\% & \textbf{-18.4}\% & 326336 & \textbf{-6.7}\% & -2.6\% & -1.9\% \\
		HEVC B & 306 (204+102) & -4.9\% & \textbf{-12.5}\% & \textbf{-16.8}\% & 326336 & \textbf{-5.7}\% & -1.6\% & -2.2\% \\
		HEVC C & 486 (384+102) & -3.3\% & \textbf{-12.1}\% & \textbf{-18.0}\% & 326336 & \textbf{-5.0}\% & -3.4\% & -5.0\% \\
		HEVC D & 486 (384+102) & 2.8\% & \textbf{-7.8}\% & \textbf{-16.7}\% & 326336 & \textbf{-3.8}\% & -1.7\% & -2.6\% \\
		HEVC E & 486 (384+102) & -3.4\% & \textbf{-10.2}\% & \textbf{-17.0}\% & 326336 & \textbf{-8.6}\% & -5.2\% & -5.6\% \\ \bottomrule
	\end{tabular}
\end{table*}
\begin{table*}[htb!]
	\centering
	\caption{Comparison of average BD Rate savings and complexity with a pretrained CNN approach in Low Delay P mode.}
	\label{tab:exp_ldp_comparison}
	\begin{tabular}{@{}lrrrrrrrr@{}}
		\toprule
		& \multicolumn{4}{c}{Our's} & \multicolumn{4}{c}{Jia et al. \cite{Jia2019}} \\ \midrule
		& \begin{tabular}[c]{@{}r@{}}Complexity (Y+U/V)\\ (MAC/Pix)\end{tabular} & Y & U & V & \begin{tabular}[c]{@{}r@{}}Complexity\\ (MAC/Pix)\end{tabular} & Y & U & V \\ \midrule
		HEVC A & 216 (114+102) & \textbf{-9.1}\% & \textbf{-15.7}\% & \textbf{-19.5}\% & 326336 & -3.5\% & 0.2\% & 0.3\% \\
		HEVC B & 306 (204+102) & \textbf{-6.8}\% & \textbf{-13.9}\% & \textbf{-17.8}\% & 326336 & -4.5\% & -0.5\% & -1.1\% \\
		HEVC C & 486 (384+102) & -3.2\% & \textbf{-12.4}\% & \textbf{-18.8}\% & 326336 & \textbf{-4.4}\% & -1.0\% & -3.0\% \\
		HEVC D & 486 (384+102) & 3.2\% & \textbf{-7.9}\% & \textbf{-15.5}\% & 326336 & \textbf{-3.5}\% & -0.8\% & -0.9\% \\
		HEVC E & 486 (384+102) & -5.3\% & \textbf{-12.6}\% & \textbf{-19.7}\% & 326336 & \textbf{-7.7}\% & -1.7\% & -0.9\% \\ \bottomrule
	\end{tabular}
\end{table*}
\subsection{Comparison to pretrained CNNs}
The previous sections presented and analysed our results under different configurations with varying parameter settings. In this section, we compare our results to the CNN-based denoising approach of Jia et al. \cite{Jia2019}, who propose an ensemble of networks. An additional discrimination network is responsible to chose the network to denoise a particular patch. Jia et al. \cite{Jia2019} showed that they outperform similar approaches like VRCNN \cite{Dai2017b} and VDSR \cite{Kim2016a} and at the same time put an emphasis on parameter and complexity reduction in their network design. This offers a good comparison to our low complexity online learning approach.

Table~\ref{tbl:exp_ra_comp} compares the two approaches in the random access setting. In accordance with the results presented above, we chose 2/2 and 1/2 pixel packing for the luma channel of HEVC A and B, respectively. All remaining results are obtained without pixel packing. Despite having about three orders of magnitude less complexity, our algorithm performs well on par for HEVC A and D luma, and underperforms by 0.6\% and 0.8\% on the luma of HEVC B and C, respectively. For HEVC E, though, our algorithm is clearly outperformed by the static CNN. This may be by the low bit rate for HEVC E sequences and the fact that their content contains very little dynamic, where error recovery by a statically trained neural network may be easier than in scenes with a low of motion. For chroma channels, on the other hand, our algorithm performs favourably on all test sets.

In the low delay setting, results differ between bi-directional (Table~\ref{tab:exp_ldb_comparison}) and uni-directional (Table~\ref{tab:exp_ldp_comparison}) prediction. As discussed before, our algorithm performs more efficient in high bit rate (high resolution) settings, hence we perform favourably in the LDP setting of HEVC A and B and contain the shortfall to 0.8\% in LDB. In low bit rate settings (HEVC C-E), our algorithm is outperformed on the Y channel, while our chroma gains remain significantly above those of Jia et al. for all test sets.

Overall, this shows that our algorithm can perform at least on par with pretrained CNNs in high resolution settings and reduce the computational cost at the decoder to 0.1\% of that of a pretrained CNN.
\begin{table*}[htb!]
	\centering
	\caption{
		\textcolor{black}{Comparison of average BD Rate savings for the luma component when ALF is active (first and second group of columns) and our results w/o ALF (third group) and ALF-only results (fourth group) for reference. Complexities for our method are as shown in Tables~\ref{tbl:exp_ra_comparison}, \ref{tab:exp_ldb_comparison}, and \ref{tab:exp_ldp_comparison}. DB and SAO are enabled for all experiments.}}
	\label{tab:exp_alf_comparison}
	\begin{tabular}{@{}lrrrrrrrrrrrrrrr@{}}
		\toprule
		Configuration & 
		  \multicolumn{3}{c}{HEVC + ALF + Jia et al. \cite{Jia2019}} & & \multicolumn{3}{c}{HEVC + ALF + Our's} & & \multicolumn{3}{c}{HEVC + Our's} & &
		  \multicolumn{3}{c}{HEVC + ALF}\\ \midrule
		Anchor & 
		  \multicolumn{3}{c}{HEVC + ALF} & & \multicolumn{3}{c}{HEVC + ALF} & & \multicolumn{3}{c}{HEVC} & & \multicolumn{3}{c}{HEVC} \\ 
		  \midrule
		&  RA & LDB & LDP & & RA & LDB & LDP & & RA & LDB & LDP & & RA & LDB & LDP \\ \midrule
		HEVC A  
		& -3.1\% & -3.2\% & -3.2\% & & -4.4\% & -4.0\% & -4.1\% & & -6.8\% & -5.9\% & -9.1\% & & -4.8\% & -4.9\% & -7.1\%  \\
		HEVC B 
		& -2.7\% & -2.5\% & -2.9\% & & -4.1\% & -2.7\% & -3.4\% & & -5.9\% & -4.9\% & -6.8\% & & -3.4\% & -3.0\% & -6.2\% \\
		HEVC C 
		 & -3.7\% & -4.0\% & -3.8\% & & -2.2\% & -1.7\% & -1.6\% & & -3.7\% & -3.3\% & -3.2\% & & -2.1\% & -1.8\% & -2.3\% \\
		HEVC D 
		& -3.4\% & -3.4\% & -3.7\% & & -1.5\% & 3.6\% & 3.3\% & & -3.4\% & 2.8\% & 3.2\% & & -2.2\% & -1.4\% & -0.7\% \\
		HEVC E 
		& -5.3\% & -5.8\% & -6.2\% & & -1.6\% & -0.4\% & -1.4\% & & -2.8\% & -3.4\% & -5.3\% & & -3.0\% & -2.7\% & -4.0\%\\ \bottomrule
	\end{tabular}
\end{table*}
\subsection{Combination with ALF}
\textcolor{black}{
Adaptive Loop Filtering (ALF) \cite{Tsai2013a} has previously been proposed as an in-loop-filter coding tool to reduce distortion based on a filter that is adapted frame-wise.  Table~\ref{tab:exp_alf_comparison} shows the BD rate savings of our method when applied to a codec with ALF, DB, and SAO active (second group of columns). The rate savings are calculated with respect to an anchor with ALF, DB and SAO. We only report the luma component as this is what ALF is applied to. When comparing the improvement over ALF to the results reported without ALF activated (third group of columns), one notices a drop in coding gain across all test sets and modes. Jia et al.'s \cite{Jia2019} observed the same phenomenon for their pretrained CNN method. Both methods have an overlap with ALF's function of reducing noise by linear filtering, hence such an influence is not unexpected. 
\\
Comparing how our method performs over ALF with Jia et al.'s \cite{Jia2019} method over ALF (columns groups one and two), their pretrained CNN performs favourably on the luma component for lower bit-rate scenarios. When the overhead generated by our method is less significant, as is the case for HEVC A \& B, our method outperforms the far more complex pretrained CNN. This shows that, especially for high-resolution video coding, our method can provide additional coding gain over Adaptive Loop Filtering. 
\\
Directly comparing how our method improves HEVC with the improvements from ALF (column groups three and four), that except for HEVC D low delay and HEVC E random access modes, our method outperforms ALF across all data sets and coding modes. With coding gain differences of up to 2.5\% for HEVC B in random access mode, our method overall performs favourably against ALF in direct comparison.}
\section{Complexity}
\label{sec:complexity}
Complexity is a major issue in video coding, especially for the decoder. By design, our approach is asymmetric, requiring a higher complexity at the encoder but enabling a lower complexity at the decoder. 
\begin{table*}[ht!]
\centering
\caption{Encoding complexity of our algorithm relative to the HM-16.17 baseline and absolute w/ network complexities as shown in tables \ref{tbl:exp_ra_comparison}, \ref{tab:exp_ldb_comparison}, and \ref{tab:exp_ldp_comparison}. Low Delay with a GoP of 5 is used. The relative timing is computed by dividing the runtime of HM-16.17 \textit{including} our algorithm by the original runtime of HM-16.17.}
\label{tab:encoding_complexity}
\begin{tabular}{@{}lrrrrrrrr@{}}
\toprule
Device & \multicolumn{4}{c}{CPU} & \multicolumn{4}{c}{GPU} \\ \midrule
Setting & \multicolumn{1}{l}{RA (32 Frames)} & \multicolumn{1}{l}{RA (256 Frames)} & \multicolumn{1}{l}{LDB} & \multicolumn{1}{l}{LDP} & \multicolumn{1}{l}{RA (32 Frames)} & \multicolumn{1}{l}{RA (256 Frames)} & \multicolumn{1}{l}{LDB} & \multicolumn{1}{l}{LDP} \\ \midrule
HEVC A & 124\% & 104\% & 113\% & 117\% & 101\% & 101\% & 101\% & 101\% \\
HEVC B & 168\% & 109\% & 136\% & 143\% & 103\% & 101\% & 102\% & 102\% \\
HEVC C & 434\% & 142\% & 295\% & 346\% & 113\% & 102\% & 108\% & 111\% \\
HEVC D & 1659\% & 295\% & 937\% & 1144\% & 163\% & 108\% & 136\% & 145\% \\
HEVC E & 714\% & 178\% & 416\% & 506\% & 125\% & 104\% & 114\% & 117\% \\ \midrule
Average & 620\% & 166\% & 379\% & 451\% & 121\% & 104\% & 112\% & 115\% \\ \midrule
\begin{tabular}[c]{@{}l@{}}Jia et al.~\cite{Jia2019}\\(Avg. over RA/LDB/LDP)\end{tabular}
 & \multicolumn{4}{c}{} & \multicolumn{4}{c}{213\%} \\ \midrule
 \multicolumn{9}{c}{Absolute Encoding Time (CNN only; seconds/frame)} \\\midrule
 HEVC A & 18.6 & 2.3 & 12.4 & 12.4 & 0.76 & 0.1 & 0.6 & 0.6 \\
 HEVC B & 22.3 & 2.8 & 15.6 & 15.6 & 0.86 & 0.11 & 0.66 & 0.66 \\
 HEVC C/D/E & 27.1 & 3.4 & 19.6 & 19.6 & 1.1 & 0.14 & 0.84& 0.84 \\
  \bottomrule
\end{tabular}
\end{table*}
Table~\ref{tab:encoding_complexity} shows the encoding complexity relative to the HM-16.17 baseline as $\frac{T'}{T}$ where $T'$ is the total runtime of HM-16.17 and our algorithm and $T$ is the original HM runtime. \textcolor{black}{The absolute encoding times are shown as well. Note that the network's training time is not changed by the resolution as the hyperparameters stay constant. Hence, as the network complexities for HEVC C/D/E are identical, their encoding times are equal as well. While real time encoding is not yet possible, this shows that online training can be feasible.} Comparing the two platforms, the CPU (Intel i7-3770 @ 3.40GHz) performs significantly slower than the GPU. Besides the GPU's parallel processing capabilities, the used CPU is about 6 years old and more recent CPUs have even better support for SIMD operations. Jia et al.~\cite{Jia2019} note that their encoding (using GPU) takes on average 213\% (they give an overhead of 113\% \textit{without} the HM-16.17 running time). Comparing that to our averages, ranging from 112\% to 121\%, our approach encodes significantly faster even though the network is trained during encoding while Jia et al. have a pretrained ensemble available. It should further be noted, that we did not tune hyperparamers like the number of iterations or batch size to optimize the training times. In addition, when we apply our algorithm to longer RAS (up to 256 frames) as reported in Table~\ref{tbl:exp_ra_ras_size}, the encoding time of our proposed algorithm remains constant as neither the batch size nor the number of iterations need to be adjusted. This reduces the overall overhead even further. While such a configuration is not part of the HEVC Common Test Conditions evaluation settings, it's often used in practice.\\
\begin{table}[ht!]
	\centering
	\caption{Decoding complexity of our algorithm relative to the HM-16.17 baseline. Low Delay with a GoP of 5 is used. The relative timing is computed by dividing the runtime of HM-16.17 \textit{including} our algorithm by the original runtime of HM-16.17.}
	\label{tab:decoding_complexity}
	\begin{tabular}{@{}lrrrrrr@{}}
		\toprule
		Device & \multicolumn{3}{c}{CPU} & \multicolumn{3}{c}{GPU} \\ \midrule
		Setting & \multicolumn{1}{l}{RA}  & \multicolumn{1}{l}{LDB} & \multicolumn{1}{l}{LDP} & \multicolumn{1}{l}{RA}  & \multicolumn{1}{l}{LDB} & \multicolumn{1}{l}{LDP} \\ \midrule
		HEVC A & 417\% & 381\% & 356\% & 128\% & 125\% & 123\%  \\
		HEVC B & 416\% & 393\% & 423\% & 129\% & 128\% & 130\%  \\
		HEVC C & 488\% & 449\% & 467\% & 154\% & 149\% & 151\%  \\
		HEVC D & 456\% & 436\% & 455\% & 209\% & 203\% & 203\% \\
		HEVC E & 783\% & 906\% & 932\% & 192\% & 204\% & 206\% \\ \midrule
		Average & 512\% & 514\% & 527\% & 162\% & 162\% & 163\% \\ \midrule
		\begin{tabular}[c]{@{}l@{}}Jia et al.~\cite{Jia2019}\\(Avg.)\end{tabular}
		& \multicolumn{3}{c}{} & \multicolumn{3}{c}{11756\%} \\ \bottomrule
	\end{tabular}
\end{table}
Table~\ref{tab:decoding_complexity} shows the decoding complexity of our algorithm for different devices and settings. While the overhead is generally higher than for the encoding case, the CPU, for the reasons noted above, runs once again significantly slower. As the decoding process is similar for random access and low delay modes, their is little difference between their respective time complexities. When compared to the application of a pretrained CNN, the advantage of our algorithm becomes even more obvious. At the expense of training the parameters at encoding time, our approach yields a network significantly smaller than a pretrained alternative, reducing decoding complexity by several orders of magnitude. \textcolor{black}{A pretrained CNN needs to cope with all possible distributions of decoded and residual signals, not only those present in a short sequence. As a modern codec like HEVC may yield very complex error signals, a single model needs to be sufficiently expressive. For deep learning, this means a model requires more layer with more filters, which increases the number of operations required for computation. Our approach of a tiny CNN tailored to a series of frames with probably similar content on the other hand can be advantageous in circumventing this problem.} 
\textcolor{black}{Complexity-wise, noting from Tables~\ref{tbl:exp_ra_comparison}, \ref{tab:exp_ldb_comparison}, and \ref{tab:exp_ldp_comparison} that Jia et al.'s \cite{Jia2019} network has more than 500 times the complexity of our method, one could expect their average decoding time to be even higher (about $500\cdot 62\%=31000\%$) when scaling up our 62\% overhead by the complexity ratio. However, their network has more filters and thereby a higher degree of parallelism. This enables their algorithm a higher GPU utilization and therefore a better up-scaling to only 11756\%. Future GPU implementations of convolutional layers may improve the GPU utilization for smaller networks, which could further improve the runtime of our algorithm.}
\section{Conclusion}
\label{sec:conclusion}
In this paper, we propose an online learning algorithm to exploit non-local redundancies in High Efficiency Video Coding. The novelty of our approach resides in the ability to learn parameters at encoding time and transmit those to the decoder in order to enable low complexity non-linear denoising. We propose a network design that is efficient enough for both PSNR improvement and signalling as part of the video bit stream. Extensive experimental results shed light on certain aspects of our algorithm design and demonstrate favourable performance over the HEVC CTC baseline and, for high resolutions, over a statically trained CNN ensemble in terms of coding gain. The low complexity design makes practical applications possible and thereby increases the potential impact of this work on future video coding technologies.

\bibliographystyle{IEEEtran}
\bibliography{main.bib}

\begin{thebibliography}{10}
\providecommand{\url}[1]{#1}
\csname url@samestyle\endcsname
\providecommand{\newblock}{\relax}
\providecommand{\bibinfo}[2]{#2}
\providecommand{\BIBentrySTDinterwordspacing}{\spaceskip=0pt\relax}
\providecommand{\BIBentryALTinterwordstretchfactor}{4}
\providecommand{\BIBentryALTinterwordspacing}{\spaceskip=\fontdimen2\font plus
\BIBentryALTinterwordstretchfactor\fontdimen3\font minus
  \fontdimen4\font\relax}
\providecommand{\BIBforeignlanguage}[2]{{%
\expandafter\ifx\csname l@#1\endcsname\relax
\typeout{** WARNING: IEEEtran.bst: No hyphenation pattern has been}%
\typeout{** loaded for the language `#1'. Using the pattern for}%
\typeout{** the default language instead.}%
\else
\language=\csname l@#1\endcsname
\fi
#2}}
\providecommand{\BIBdecl}{\relax}
\BIBdecl

\bibitem{Wu2018}
C.~Y. Wu, N.~Singhal, and P.~Kr{\"{a}}henb{\"{u}}hl, ``{Video compression
  through image interpolation},'' vol. 11212 LNCS, apr 2018.

\bibitem{Minnen2018}
D.~Minnen, J.~Ball{\'{e}}, and G.~Toderici, ``{Joint Autoregressive and
  Hierarchical Priors for Learned Image Compression},'' in \emph{Neural
  Information Processing Systems}, 2018.

\bibitem{Liu2019}
H.~Liu, T.~Chen, P.~Guo, Q.~Shen, and Z.~Ma, ``{Gated Context Model with
  Embedded Priors for Deep Image Compression},'' feb 2019.

\bibitem{Mentzer2018}
F.~Mentzer, E.~Agustsson, M.~Tschannen, R.~Timofte, and L.~{Van Gool},
  ``{Conditional Probability Models for Deep Image Compression},'' 2018.

\bibitem{Theis2017}
L.~Theis, W.~Shi, A.~Cunningham, and F.~Husz{\'{a}}r, ``{Lossy Image
  Compression with Compressive Autoencoders},'' \emph{ICLR}, 2017.

\bibitem{Toderici2016}
G.~Toderici, D.~Vincent, N.~Johnston, S.~J. Hwang, D.~Minnen, J.~Shor, and
  M.~Covell, ``{Full Resolution Image Compression with Recurrent Neural
  Networks},'' \emph{Computer Vision and Pattern Recognition}, 2016.

\bibitem{Johnston2017}
N.~Johnston, D.~Vincent, D.~Minnen, M.~Covell, S.~Singh, T.~Chinen, S.~J.
  Hwang, J.~Shor, and G.~Toderici, ``{Improved Lossy Image Compression with
  Priming and Spatially Adaptive Bit Rates for Recurrent Networks},''
  \emph{Computer Vision and Pattern Recognition}, 2017.

\bibitem{Rippel2017}
O.~Rippel and L.~Bourdev, ``{Real-Time Adaptive Image Compression},''
  \emph{ICML}, 2017.

\bibitem{Balle2017}
J.~Ball{\'{e}}, V.~Laparra, and E.~P. Simoncelli, ``{End-to-end Optimized Image
  Compression},'' \emph{ICLR}, 2017.

\bibitem{Li2017}
M.~Li, W.~Zuo, S.~Gu, D.~Zhao, and D.~Zhang, ``{Learning Convolutional Networks
  for Content-weighted Image Compression},'' 2017.

\bibitem{Baig2017}
M.~H. Baig, V.~Koltun, and L.~Torresani, ``{Learning to Inpaint for Image
  Compression},'' \emph{NIPS}, 2017.

\bibitem{Toderici2015}
G.~Toderici, S.~M. O'Malley, S.~J. Hwang, D.~Vincent, D.~Minnen, S.~Baluja,
  M.~Covell, and R.~Sukthankar, ``{Variable Rate Image Compression with
  Recurrent Neural Networks},'' \emph{International Conference On Learning
  Representations}, 2015.

\bibitem{Agustsson2017}
E.~Agustsson, F.~Mentzer, M.~Tschannen, L.~Cavigelli, R.~Timofte, L.~Benini,
  and L.~{Van Gool}, ``{Soft-to-Hard Vector Quantization for End-to-End
  Learning Compressible Representations},'' \emph{NIPS}, 2017.

\bibitem{Balle2018}
J.~Ball{\'{e}}, D.~Minnen, S.~Singh, S.~J. Hwang, and N.~Johnston,
  ``{Variational image compression with a scale hyperprior},''
  \emph{International Conference On Learning Representations}, 2018.

\bibitem{Klopp}
J.~P. Klopp, Y.-c.~F. Wang, and L.-g. Chen, ``{Learning a Code-Space Predictor
  by Exploiting Intra-Image-Dependencies Review of Learned Image
  Compression},'' in \emph{British Machine Vision Conference}, 2018.

\bibitem{Wiegand2003}
T.~Wiegand, G.~J. Sullivan, G.~Bj{\o}ntegaard, and A.~Luthra, ``{Overview of
  the H.264/AVC video coding standard},'' \emph{IEEE Transactions on Circuits
  and Systems for Video Technology}, 2003.

\bibitem{Sullivan2012a}
G.~J. Sullivan, J.~R. Ohm, W.~J. Han, and T.~Wiegand, ``{Overview of the high
  efficiency video coding (HEVC) standard},'' \emph{IEEE Transactions on
  Circuits and Systems for Video Technology}, 2012.

\bibitem{Kim1999}
S.~D. Kim, J.~Yi, H.~M. Kim, and J.~B. Ra, ``{A deblocking filter with two
  separate modes in block-based video coding},'' \emph{IEEE Transactions on
  Circuits and Systems for Video Technology}, 1999.

\bibitem{List2003}
P.~List, A.~Joch, J.~Lainema, G.~Bj{\o}ntegaard, and M.~Karczewicz, ``{Adaptive
  deblocking filter},'' \emph{IEEE Transactions on Circuits and Systems for
  Video Technology}, 2003.

\bibitem{Norkin2012}
A.~Norkin, G.~Bj{\o}ntegaard, A.~Fuldseth, M.~Narroschke, M.~Ikeda,
  K.~Andersson, M.~Zhou, and G.~{Van Der Auwera}, ``{HEVC deblocking filter},''
  \emph{IEEE Transactions on Circuits and Systems for Video Technology}, 2012.

\bibitem{Jo2016}
H.~Jo, S.~Park, and D.~Sim, ``{Parallelized deblocking filtering of HEVC
  decoders based on complexity estimation},'' \emph{Journal of Real-Time Image
  Processing}, 2016.

\bibitem{Fu2012}
C.~M. Fu, E.~Alshina, A.~Alshin, Y.~W. Huang, C.~Y. Chen, C.~Y. Tsai, C.~W.
  Hsu, S.~M. Lei, J.~H. Park, and W.~J. Han, ``{Sample adaptive offset in the
  HEVC standard},'' \emph{IEEE Transactions on Circuits and Systems for Video
  Technology}, 2012.

\bibitem{Tsai2013a}
C.~Y. Tsai, C.~Y. Chen, T.~Yamakage, I.~S. Chong, Y.~W. Huang, C.~M. Fu,
  T.~Itoh, T.~Watanabe, T.~Chujoh, M.~Karczewicz, and S.~M. Lei, ``{Adaptive
  loop filtering for video coding},'' \emph{IEEE Journal on Selected Topics in
  Signal Processing}, 2013.

\bibitem{Zhang2017}
X.~Zhang, R.~Xiong, W.~Lin, J.~Zhang, S.~Wang, S.~Ma, and W.~Gao,
  ``{Low-Rank-Based Nonlocal Adaptive Loop Filter for High-Efficiency Video
  Compression},'' \emph{IEEE Transactions on Circuits and Systems for Video
  Technology}, 2017.

\bibitem{Krutz2012}
A.~Krutz, A.~Glantz, M.~Tok, M.~Esche, and T.~Sikora, ``{Adaptive global motion
  temporal filtering for high efficiency video coding},'' \emph{IEEE
  Transactions on Circuits and Systems for Video Technology}, 2012.

\bibitem{Dong2015}
C.~Dong, Y.~Deng, C.~C. Loy, and X.~Tang, ``{Compression artifacts reduction by
  a deep convolutional network},'' in \emph{Proceedings of the IEEE
  International Conference on Computer Vision}, 2015.

\bibitem{Zhang2017b}
K.~Zhang, W.~Zuo, Y.~Chen, D.~Meng, and L.~Zhang, ``{Beyond a Gaussian
  denoiser: Residual learning of deep CNN for image denoising},'' \emph{IEEE
  Transactions on Image Processing}, 2017.

\bibitem{Zhang2017a}
L.~Zhang and W.~Zuo, ``{Image Restoration: From Sparse and Low-Rank Priors to
  Deep Priors [Lecture Notes]},'' \emph{IEEE Signal Processing Magazine}, 2017.

\bibitem{Zhang2017c}
K.~Zhang, W.~Zuo, S.~Gu, and L.~Zhang, ``{Learning deep CNN denoiser prior for
  image restoration},'' in \emph{Proceedings - 30th IEEE Conference on Computer
  Vision and Pattern Recognition, CVPR 2017}, 2017.

\bibitem{Yan2017}
N.~Yan, D.~Liu, H.~Li, and F.~Wu, ``{A convolutional neural network approach
  for half-pel interpolation in video coding},'' in \emph{Proceedings - IEEE
  International Symposium on Circuits and Systems}, 2017.

\bibitem{Yang2018}
R.~Yang, M.~Xu, T.~Liu, Z.~Wang, and Z.~Guan, ``{Enhancing Quality for HEVC
  Compressed Videos},'' \emph{IEEE Transactions on Circuits and Systems for
  Video Technology}, 2018.

\bibitem{Li2017a}
C.~Li, L.~Song, R.~Xie, and W.~Zhang, ``{CNN based post-processing to improve
  HEVC},'' in \emph{2017 IEEE International Conference on Image Processing
  (ICIP)}.\hskip 1em plus 0.5em minus 0.4em\relax IEEE, sep 2017.

\bibitem{Yang2017}
R.~Yang, M.~Xu, and Z.~Wang, ``{Decoder-side HEVC quality enhancement with
  scalable convolutional neural network},'' in \emph{2017 IEEE International
  Conference on Multimedia and Expo (ICME)}.\hskip 1em plus 0.5em minus
  0.4em\relax IEEE, jul 2017.

\bibitem{Cavigelli2017}
L.~Cavigelli, P.~Hager, and L.~Benini, ``{CAS-CNN: A deep convolutional neural
  network for image compression artifact suppression},'' in \emph{2017
  International Joint Conference on Neural Networks (IJCNN)}.\hskip 1em plus
  0.5em minus 0.4em\relax IEEE, may 2017.

\bibitem{Dai2017}
Y.~Dai, D.~Liu, and F.~Wu, ``{A convolutional neural network approach for
  post-processing in HEVC intra coding},'' 2017.

\bibitem{Zhang2018}
Y.~Zhang, T.~Shen, X.~Ji, Y.~Zhang, R.~Xiong, and Q.~Dai, ``{Residual Highway
  Convolutional Neural Networks for in-loop Filtering in HEVC},'' \emph{IEEE
  Transactions on Image Processing}, vol.~27, no.~8, 2018.

\bibitem{Jia2019}
C.~Jia, S.~Wang, X.~Zhang, S.~Wang, J.~Liu, S.~Pu, and S.~Ma, ``{Content-Aware
  Convolutional Neural Network for In-loop Filtering in High Efficiency Video
  Coding},'' \emph{IEEE Transactions on Image Processing}, jan 2019.

\bibitem{Lu2019}
G.~Lu, W.~Ouyang, D.~Xu, X.~Zhang, C.~Cai, and Z.~Gao, ``{DVC: An End-to-end
  Deep Video Compression Framework},'' in \emph{Computer Vision and Patter
  Recognition}, nov 2019.

\bibitem{Rippel2018}
O.~Rippel, S.~Nair, C.~Lew, S.~Branson, A.~G. Anderson, and L.~Bourdev,
  ``{Learned Video Compression},'' 2018.

\bibitem{Howard2017}
A.~G. Howard, M.~Zhu, B.~Chen, D.~Kalenichenko, W.~Wang, T.~Weyand,
  M.~Andreetto, and H.~Adam, ``{MobileNets},'' \emph{arXiv preprint
  arXiv:1704.04861}, 2017.

\bibitem{Ohm2012}
J.~R. Ohm, G.~J. Sullivan, H.~Schwarz, T.~K. Tan, and T.~Wiegand, ``{Comparison
  of the coding efficiency of video coding standards-including high efficiency
  video coding (HEVC)},'' \emph{IEEE Transactions on Circuits and Systems for
  Video Technology}, 2012.

\bibitem{Ioffe2015}
S.~Ioffe and C.~Szegedy, ``{Batch Normalization: Accelerating Deep Network
  Training by Reducing Internal Covariate Shift},'' \emph{International
  Conference on Machine Learning}, feb 2015.

\bibitem{Chih2018}
C.~Y. Chih, S.~S. Wu, J.~P. Klopp, and L.~G. Chen, ``{Accurate and Bandwidth
  Efficient Architecture for CNN-based Full-HD Super-Resolution},'' in
  \emph{Proceedings - IEEE International Symposium on Circuits and Systems},
  2018.

\bibitem{Courbariaux2016}
M.~Courbariaux and Y.~Bengio, ``{BinaryNet: Training Deep Neural Networks with
  Weights and Activations Constrained to +1 or -1},'' \emph{arXiv}, 2016.

\bibitem{Rastegari2016}
M.~Rastegari, V.~Ordonez, J.~Redmon, and A.~Farhadi, ``{XNOR-Net: ImageNet
  Classification Using Binary Convolutional Neural Networks},'' \emph{arXiv},
  2016.

\bibitem{Paszke2017}
A.~Paszke, S.~Gross, S.~Chintala, G.~Chanan, E.~Yang, Z.~DeVito, Z.~Lin,
  A.~Desmaison, L.~Antiga, and A.~Lerer, ``{Automatic Differentiation in
  {\{}PyTorch{\}}},'' in \emph{NIPS Autodiff Workshop}, 2017.

\bibitem{NIPS2012_4522}
J.~Snoek, H.~Larochelle, and R.~P. Adams, ``{Practical Bayesian Optimization of
  Machine Learning Algorithms},'' in \emph{Advances in Neural Information
  Processing Systems 25}, F.~Pereira, C.~J.~C. Burges, L.~Bottou, and K.~Q.
  Weinberger, Eds.\hskip 1em plus 0.5em minus 0.4em\relax Curran Associates,
  Inc., 2012.

\bibitem{Kingma2015}
D.~P. Kingma and J.~L. Ba, ``{Adam: a Method for Stochastic Optimization},''
  \emph{International Conference on Learning Representations 2015}, 2015.

\bibitem{Bjontegaard2001}
G.~Bj{\o}ntegaard, ``{Calculation of Average PSNR Differences between RD
  curves. ITU-T SG16/Q6},'' ITU-T SG16/Q6, Austin, Texas, USA, Tech. Rep.,
  2001.

\bibitem{Derf2020}
\BIBentryALTinterwordspacing
Derf, ``{Xiph.org :: Derf's Test Media Collection},'' 2020. [Online].
  Available: \url{https://media.xiph.org/video/derf/}
\BIBentrySTDinterwordspacing

\bibitem{Dai2017b}
Y.~Dai, D.~Liu, and F.~Wu, ``{A Convolutional Neural Network Approach for
  Post-Processing in HEVC Intra Coding},'' in \emph{International Conference on
  Multimedia Modeling}.\hskip 1em plus 0.5em minus 0.4em\relax Springer, Cham,
  2017.

\bibitem{Kim2016a}
J.~Kim, J.~K. Lee, and K.~M. Lee, ``{Accurate image super-resolution using very
  deep convolutional networks},'' in \emph{Proceedings of the IEEE Computer
  Society Conference on Computer Vision and Pattern Recognition}, 2016.

\end{thebibliography}
\end{document}